# An equilibrium model for matching impatient demand and patient supply over time

Garud Iyengar [*]   Anuj Kumar [†]

October 13, 2018

**Abstract**

We present a simple dynamic equilibrium model for an online exchange where both buyers and sellers arrive according to a exogenously defined stochastic process. The structure of this exchange is motivated by the limit order book mechanism used in stock markets. Both buyers and sellers are elastic in the price-quantity space; however, only the sellers are assumed to be patient, i.e. only the sellers have a price - time elasticity, whereas the buyers are assumed to be impatient. Sellers select their selling price as a best response to all the other sellers' strategies. We define and establish the existence of the equilibrium in this model and show how to numerically compute this equilibrium. We also show how to compute other relevant quantities such as the equilibrium expected time to sale and equilibrium expected order density, as well as the expected order density conditioned on current selling price. We derive a closed form for the equilibrium distribution when the demand is price independent. At this equilibrium the selling (limit order) price distribution is power tailed as is empirically observed in order driven financial markets.

## 1 Introduction

Standard models for dynamic economies have had limited success in predicting the real world market dynamics mainly because of the following two reasons: first, it is difficult to analyze a dynamic market with complex dynamics, consequently models that are solvable are necessarily simple abstractions that are not able to incorporate many important features of "real" markets; second, in order to remain tractable economic models need to assume that agents are rational; however, rationality imposes such unreasonable demands that agents are almost never rational in "real" markets.

To overcome these issues, many models (see, e.g. Daniels et al. (2003); Farmer et al. (2005); Farmer and Zovko (2002); Luckock (2003)) for dynamic markets assume that the actions of the agents are randomly distributed according to a distribution that is chosen to reflect nominal economic behavior of agents. In such models, the interesting features of the statistical behavior of the market is a consequence of the market dynamics itself.

[*]Industrial Engineering and Operations Research Department, Columbia University, New York, NY-10027. Email: gi10@columbia.edu

[†]Industrial Engineering and Operations Research Department, Columbia University, New York, NY-10027. Email: ak2108@columbia.edu

In this paper, we present a model that incorporates features of both of these approaches. In our model, the agents are strategic, i.e. they do not take random actions; however, agents have bounded rationality, and therefore, do not base their actions on the detailed market conditions at the arrival epoch but on the average long term market characteristics arising from the random actions of other agents. In effect, we define an equilibrium concept on the *distribution* induced by the random agent actions. In this equilibrium the long run the essential market information filters to the agents and gets reflected in the statistical properties of the market. The contributions of this paper can be summarized as follows.

(a) We define an equilibrium concept on the supply in a dynamic exchange where agents react to the long-term average impact of the actions of all the agents in the exchange. We show that such an equilibrium always exists in our model and show how to numerically compute it. We also provide closed form expressions for various statistical properties of the market, such as the expected time to execution, the average outstanding inventory and the average outstanding inventory conditional on current lowest selling price. Our model is a good approximation to online used book market at *amazon.com*, where multiple copies of substitutable products are available at different prices and sellers strategically post their selling prices based on their beliefs about the execution time.

(b) We show that our proposed equilibrium can be computed in closed form in a market where buyers are not price sensitive. In this solution, the trade execution prices exhibit power tails distribution that matches the empirically observed distribution limit-book driven financial markets.

There is a large body of literature investigating financial markets with the limit order book as a market clearing mechanism. Farmer and Zovko Farmer and Zovko (2002) demonstrate a striking regularity in the way people place limit orders in financial markets using a data set from the London Stock Exchange. They define the relative limit price as the difference between the limit price and the best price available for instantaneous execution. They conclude that for both buy and sell orders, the unconditional cumulative distribution of relative limit prices decays roughly as a power law with an exponent approximately $\beta \approx 1.5$, i.e. $\mathbb{P}(\text{limit price} > x) = \frac{A}{(x_0+x)^\beta}$. This behavior spans more than two decades, ranging from a few ticks to about 2000 ticks. Bouchaud et al Bouchaud et al. (2002) also report a power law distribution of prices. Our model shows that this type of power tail distribution can arise in equilibrium for a limit order book mechanism when more patient trader get better price in order for less patient trader to get better execution time.

The models in Luckock (2003); Mendelson (1982); Domowitz and Wang (1994) are not strategic in that they assume a stationary order arrival independent of the state of the book. This can be reasonable in very fast highly liquid market, where for small time horizon, large number of traders participate in the trading process; thus, their strategies average out to result in a random stationary behavior. In addition, these models assume that the order arrival pattern is *exogenous* to the model; consequently, their approach reduces to performance evaluation for a given supply and demand function. This approach leads to many pathologies, such as accumulation of orders outside the active window Luckock (2003). In contrast, in our model the stationary equilibrium behavior of traders gives rise to *endogenous* limit order distribution. Our model is similar to the separable markets discussed in Luckock (2003).

## 2 Continuous Online Exchange

We consider a continuous time online exchange market for a single commodity or a set of substitutable commodities. There is no fee for using the online exchange and orders cannot be canceled. This exchange only allows the posted prices to be multiples of a tick size $\epsilon$. We assume that there exists a high enough constant price $(N+1)\epsilon$ at which an exogenous seller is willing to sell unlimited amount of the commodity and, symmetrically, a low enough price (normalized to 0) at which an exogenous buyer is willing to buy an unlimited quantity. Thus, the set of allowed selling prices is a grid $\{\epsilon, 2\epsilon, \ldots, N\epsilon\}$.

The sellers on this exchange arrive according to a Poisson process with rate $\lambda$. Each seller is offering only one unit of the commodity for sale. The sellers are assumed to be risk neutral and incur a cost that is proportional to their execution time, i.e. the time elapsed between their arrival and sale of their item. On arrival, the sellers choose a selling price, or equivalently a price tick $j$, based on their beliefs about expected time to trade execution at each price tick. We assume that sellers believe that the expected time to execution depends *only* on the selling price; thus, excluding the possibility of sellers exploiting the complete state of the online exchange, specifically the number of unsold items at each tick. Note that we are implicitly assuming that sellers only react to the *long-term average impact* of the actions of all the agents in the exchange

The buyers on this exchange arrive according to an independent Poisson process with rate $\mu$. On arrival, each buyer independently decides to buys one unit with probability $\beta(p)$ when the current lowest (outstanding) selling price is $p$, where the demand function $\beta(p)$ is downward slopping, i.e. the $\beta(p)$ function is non-increasing in $p$. Buyers who arrive to find no sellers are lost. Note that our model assumes that the sellers are *patient* whereas the buyers are *impatient*.

The expected profit $u_n$ of the $n$-th arriving seller is given by

$$u_n(j_n, \delta_n) = j_n \epsilon - \delta_n T_n(j_n) \qquad (1)$$

where $j_n$ denotes the price-tick selected by the seller, $\delta_n$ is the *patience* parameter of the seller, and $T_n(j_n)$ is the expected execution time at price tick $j_n$. We have normalized the cost (or exogenous value) of the commodity to the seller to zero. The patience parameters $\{\delta_n : n \geq 1\}$ are assumed to be IID sampled from a distribution function $F_\delta$, which is assumed to be continuous with support $[0, \bar{\delta}]$. Thus, the sellers are heterogeneous in their patience parameter. We assume that this heterogeneity in patience is a manifestation of the heterogeneity in the agents' own business models. The utility function in (1) is motivated by the money value of time (e.g. lost labor, cost of tracking the trade etc) rather than time value of money (i.e. delayed payments), see Foucault et al. (2005), but we expect results similar to presented in the paper to hold for a large class of utility function that are monotone in price and execution time.

In this paper, we ask what equilibrium supply function would result from the interaction of sellers heterogenous in their patience and impatient buyers in a stationary online exchange market environment. In particular, we are interested in how this supply function depends on the market parameters such that demand elasticity and traffic intensity.

## 2.1 Information Structure

We assume that market mechanism is common knowledge i.e. all sellers know that a sequence of sellers would arrive according to a Poisson process and offer their unit of supply at price that optimizes their own utility function given by (1). The patience parameter distribution $F_\delta$ is common knowledge among the sellers, whereas the patience parameter $\delta_n$ is a private information of seller $n$.

Sellers form beliefs about the expected execution time $T_n(j)$ at the price tick $j$ based on the their information set and post their unit at the price

$$\sigma_n(\delta) = \underset{j \in \{1,2...,N\}}{\operatorname{argmax}} \{j\epsilon - \delta T_n^{\sigma_{-n}}(j)\}, \qquad (2)$$

where the notation $T_n^{\sigma_{-n}}(j)$ indicates that the belief about the execution time depends on the action of all other sellers. Note that we assume that the sellers beliefs about the expected execution time depends *only* on the price tick $j$.

Since the information sets are symmetric, we restrict ourselves to symmetric equilibrium strategies. Thus, the equilibrium beliefs of all sellers is symmetric, i.e. $T_n^{\sigma_{-n}}(j) = T(j)$. In rest of the paper, by equilibrium and beliefs we mean symmetric equilibrium and symmetric beliefs.

**Definition 1.** *A selling strategy $\sigma^*(.)$, that maps a patience parameter $\delta$ to a selling price tick in $\{1, 2, \ldots, N\}$, is a Bayesian Nash Equilibrium (BNE) if it solves (2) when $T^{\sigma^*}(j)$ is the stationary expected execution time given that all future trader follow the strategy $\sigma^*$.*

The execution time $T_n^{\sigma_{-n}}$ in (2) has a two-fold expectation: first, given the sample path of patience parameter $\delta_n$ of arriving sellers, it is expectation over the arrival process of the buyers; and second, there is the expectation over all sample paths of patience parameters. The latter expectation give rise to the notion of Bayesian Nash equilibrium. Since the patience parameters of sellers are assumed to be independent, it follows that the knowledge of a the draw $\delta_n$ does not give seller $n$ any information about the patience parameter of the other sellers.

## 2.2 Simple example

The following example illustrates our proposed model and the nature of results we establish later in this section.

Two sellers with one unit each of a given product arrive at time $t = 0$ into a market place where sellers are restricted to sell their product at a price $p \in \{p_1, p_2\}$, $(p_1 < p_2)$. Buyers arrive according to a Poisson process with $\mu_1$ (resp. $\mu_2$) if cheapest unit of the product is available at price $p_1$ (resp. $p_2$).

The sellers decide their selling prices as a function of their waiting cost rate $\delta$ at time 0 and then the market clears. Suppose at a symmetric equilibrium[1], a seller posts the price $p_j$, $j = 1, 2$, whenever the cost coefficient $\delta \in \mathcal{B}_j$, where $\mathcal{B}_1 \cup \mathcal{B}_2 = [0, \bar{\delta}]$. Let $\alpha_j = F_\delta(\{\mathcal{B}_1\})$ denote the equilibrium probability of selecting price $p_1$ at this equilibrium.

---
[1]Symmetric Bayesian Nash equilibrium of the one shot game with $\delta_n$ as private information to be precise.

Suppose whenever there are two sellers at a given price, each seller receives the next order with probability $\frac{1}{2}$. Then the expected waiting time if a seller selects price $p_1$ (resp. $p_2$) is $\frac{\alpha_2}{\mu_1} + \frac{3\alpha_1}{2\mu_2}$ (resp. $\frac{3\alpha_2}{2\mu_2} + \alpha_1(\frac{1}{\mu_1} + \frac{1}{\mu_2})$).

Since we assume that sellers only react to *long-term average* values, a seller would choose price $p_1$ if, and only if,

$$p_1 - \delta\left[\alpha_2 \frac{1}{\mu_1} + \alpha_1 \frac{3}{2\mu_2}\right] > p_2 - \delta\left[\alpha_2 \frac{3}{2\mu_2} + \alpha_1(\frac{1}{\mu_1} + \frac{1}{\mu_2})\right],$$

i.e.

$$\delta > \frac{\mu_1 \mu_2 (p_2 - p_1)}{\mu_1 - \mu_2 + \frac{\alpha_2}{2}(\mu_1 + \mu_2)}.$$

Since, at equilibrium, the probability of choosing price $p_2$ is $\alpha_2$, it follows that

$$\alpha_2 = F_\delta \left( \frac{\mu_1 \mu_2 (p_2 - p_1)}{\mu_1 - \mu_2 + \frac{\alpha_2}{2}(\mu_1 + \mu_2)} \right) \tag{3}$$

The solution to (3) uniquely characterizes the equilibrium.

Suppose $\delta \sim \text{unif}[0, 1]$, i.e. $F_\delta(x) = x$, for all $x \in [0, 1]$. Then at equilibrium

$$\alpha_2 = \frac{\sqrt{(\mu_1 - \mu_2)^2 + 4(\mu_1 + \mu_2)\mu_1 \mu_2 (p_2 - p_1)} - 2(\mu_1 - \mu_2)}{\mu_1 + \mu_2}$$

Note that the three factor affecting this equilibrium are: the price tick size $\epsilon = p_2 - p_1$, the demand elasticity $\mu_1 - \mu_2$ and the distribution $F_\delta$.

## 2.3 Stationary Equilibrium Outcomes

In order to characterize the BNE we need to be able to solve for the stationary behavior of the exchange for a given strategy $\sigma$. In the following we show that the exchange is a Markovian priority queuing system with closed form analytical solution for the waiting time.

Since $\{\delta_n : n \geq 1\}$ are IID, (2) implies that a selling strategy $\sigma$ *thins* the arriving sellers according to some probability mass function $\alpha_j^\sigma$, $j = 1, \ldots, N$, i.e. the arrival rate of sellers that post their unit at the price-tick $j$ is $\lambda \alpha_j^\sigma$, $j = 1, \ldots, N$.

Let $X_n(j)$ denote the inventory of outstanding orders at price tick $j$ when the $n$-th seller arrives in the exchange, and let $\mathbf{X}_n = (X_n(1), ..., X_n(N))$. The dynamics described in Section 2 implies that $\mathbf{X}_n$ is the queue length process for the Markovian preemptive priority queuing system with $N$ customer classes where the customer class $j$ has an arrival rates $\lambda \alpha_j^\sigma$ and a service rate $\mu \beta_j$. It follows that the properties of the state process $\{\mathbf{X}_n : n \geq 1\}$ are completely determined by thinning probabilities $\{\alpha_j^\sigma\}$. In order to emphasize this fact, we drop the superscript $\sigma$ and index variables by $\alpha$.

For $j = 1, \ldots, N$, define the traffic intensity in customer class $j$ as

$$\rho_j^\alpha \triangleq \frac{\lambda}{\mu} \cdot \frac{\alpha_j}{\beta_j}$$

The following lemma gives a closed-form solution for the steady state expected execution time $T_j^\alpha$ at the price tick $j$.

**Lemma 1.** *For $j = 1 \ldots N$, if $\sum_{i=1}^{j} \rho_i^\alpha < 1$, then the expected execution time at price tick $j$ is finite and is given by*

$$T^\alpha(j) = \frac{1}{\mu}\left(\frac{1}{1 - \sum_{i=1}^{j-1} \rho_i^\alpha}\right)\left[\frac{1}{\beta_j} + \frac{\sum_{i=1}^{j}(\frac{\rho_i^\alpha}{\beta_i})}{1 - \sum_{i=1}^{j} \rho_i^\alpha}\right] \quad (4)$$

*otherwise the queue at the price tick $j$ grows beyond bound and $T^\alpha(j) = \infty$.*

The proof of this Lemma can be found in Adan and Resing (2001) (pp 90, equation 9.5) Little's law implies that the long-term average expected inventory $Q_j^\alpha$ of unsold items at price tick $j$ is given by

$$Q_j^\alpha \triangleq \mathbb{E}X_\infty^\alpha(j) = \lambda \alpha_j T^\alpha(j) \quad (5)$$

where $\mathbf{X}_\infty^\alpha = \lim_{n \to \infty} \mathbf{X}_n^\alpha$ if it exists. Since $\rho_j^\alpha \geq 0$, and $\beta_j$ is decreasing in $j$, the expected execution time $T^\sigma(j)$ is non-decreasing in $j$ for all $\alpha$. Thus, the sellers face the following tradeoffs - a better execution price can only be obtained at the cost of a larger expected waiting time at the exchange.

Next, we characterize the average inventory seen by buyer or seller arriving to the exchange when the current outstanding price-tick is $j$, or equivalently, $\sum_{l=1}^{j-1} X_\infty^\alpha(l) = 0$.

**Lemma 2.** *Let $X_\infty(j)$ denote the steady state number of class $j$ customers in Markovian n-class priority queue with preemption. Then*

$$\mathbb{E}\left[\sum_{l=j}^{k} X_\infty(l) \Big| \sum_{l=1}^{j-1} X_\infty(l) = 0\right] = \frac{\rho_{jk}(1 - 2\rho_{1,j-1} + \rho_{1,j-1}^2 + \rho_{1,j-1}\rho_{jl})}{(1 - \rho_{1,j-1})^2(1 - \rho_{1,j-1} - \rho_{jl})},$$

*where $\rho_{ij} = \sum_{k=i}^{j} \frac{\lambda_k}{\mu_k}$, $\lambda_k$ is the arrival rate of class $k$ customers and $\mu_k$ is the service rate of the class $k$ customers.*

**Proof:** Since the distribution of $\sum_{l=j}^{k} X_\infty(l)$ and $\sum_{l=1}^{j-1} X_\infty(l)$ does not depend on the service discipline *within* the sets $\{1, \ldots, j-1\}$ and $\{j, \ldots, k\}$, it suffices to prove (6) for a queue with two customer classes and traffic intensities $\rho_2 = \rho_{jl}$ and $\rho_1 = \rho_{1j}$.

The generating function for the joint distribution of $(X_\infty(1), X_\infty(2))$ is given by (see (3.15) on pp. 95 in Jaiswal (1968))

$$\hat{\Pi}(z_1, z_2) = (1 - \rho_1 - \rho_2)\left[1 + \left(\frac{\lambda_1\{z_1 - \bar{b}_1[\lambda_2(1 - \alpha_2)]\}}{\lambda_1(1 - z_1) + \lambda_2(1 - z_2)}\right) \times \left(\frac{1 - \overline{S}_1\{\lambda_1(1 - z_1) + \lambda_2(1 - z_2)\}}{1 - \frac{1}{z_1}\overline{S}_1\{\lambda_1(1 - z_1) + \lambda_2(1 - z_2)\}}\right)\right]$$

$$\times \frac{(z_2 - 1)\bar{c}_2\{\lambda_2(1 - z_2)\}}{z_2 - \bar{c}_2\{\lambda_2(1 - z_2)\}}$$

where

$$\bar{b}_1(s) = \frac{\mu_1 + \lambda_1 + s - \sqrt{(\mu_1 + \lambda_1 + s)^2 - 4\lambda_1\mu_1}}{2\lambda_1},$$

$$\overline{S}_i(s) = \frac{\mu_i}{\mu_i + s} \quad i = 1, 2,$$

$$\bar{c}_2(s) = \overline{S}_2\{\lambda_1(1 - \bar{b}_1(s)) + s\}.$$

We obtain the generating function $\Pi(z)$ for distribution of $X_\infty(2) \mid X_\infty(1) = 0$ by substituting $z_1 = 0$ and $z_2 = z$ in the expression for $\hat{\Pi}(z_1, z_2)$, i.e.

$$\Pi(z) = (1 - \rho_1 - \rho) \frac{(z-1)\bar{c}_2\{\lambda_2(1-z)\}}{z - \bar{c}_2\{\lambda_2(1-z)\}}$$

The expectation $\mathbb{E}[X_\infty(2)|X_\infty(1) = 0] = \left.\frac{\partial \Pi(z)}{\partial z}\right|_{z=1}$. Simplifying we get,

$$\mathbb{E}[X_\infty(2)|X_\infty(1) = 0] = \frac{\rho_2(1 - 2\rho_1 + \rho_1^2 + \rho_1\rho_2)}{(1-\rho_1)^2(1-\rho_1-\rho_2)}$$

∎

Lemma 2 describes the steady state expected inventory the buyers or sellers see when the current price is $j\epsilon$. Although we restrict the sellers to base their beliefs of the execution time on *only* on the price tick $j$, Lemma 2 allows us to reconstruct the *entire* state of the exchange when a seller places an order.

Before characterizing the equilibrium selling strategy, we would like to point out that the distribution of the steady state price $s^\alpha$ that an arriving buyer observes is

$$\mathbb{P}(s^\alpha > j\epsilon) = 1 - \sum_{k=1}^{j} \rho_k \qquad (6)$$

Note that this distribution is not the same as the distribution of trade execution prices $\alpha_j$. The distribution $s^\alpha$ is an average over time, and latter, i.e. $\{\alpha_j\}$, is an average over orders. In this paper we will focus on the latter. Also, it is easy to notice that if the demand is inelastic, i.e. $\beta(p) \equiv 1$, the distribution of $s$ conditional on it being finite is same as the distribution $\alpha$. We consider inelastic demand in Section 3.

By Definition 1 a selling strategy $\sigma^*$ is a symmetric BNE strategy iff

$$\sigma^*(\delta) \in \underset{k \in \{1,\ldots,N\}}{\mathrm{argmax}} \{k\epsilon - \delta T^{\sigma^*}(k)\}, \qquad (7)$$

equivalently, a distribution $\alpha^*$ is a symmetric BNE strategy iff

$$\alpha_j^* = \mathbb{P}\left\{j \in \underset{1 \leq k \leq N}{\mathrm{argmax}} \left\{k\epsilon - \delta T^{\alpha^*}(k)\right\}\right\} \qquad (8)$$

The following theorem establish the existence of a symmetric equilibrium in this model.

**Theorem 1.** *An equilibrium distribution $\alpha^*$ satisfying (8) always exists. Furthermore, suppose the expected execution time $T^{\alpha^*}(j)$ is strictly increasing and convex in $j$, i.e. for $j = 2, \ldots, N$ the difference $T^{\alpha^*}(j) - T^{\alpha^*}(j-1)$ is non-negative and non-decreasing in $j$. Then every equilibrium selling strategy $\sigma^*$ that results in $\alpha^*$ is non-increasing in $\delta$ i.e. at the equilibrium $\alpha^*$ the seller with higher waiting cost rate post a lower selling price.*

**Proof:** Define $\Psi : \mathcal{S}^N \mapsto \mathcal{S}^N$ as follows

$$\Psi_j(\alpha) = \mathbb{P}\left[\delta \in \{x \in [0,\bar{\delta}] | j\epsilon - xT^\alpha(j) \geq k\epsilon - xT^\alpha(k) \; \forall k\}\right], \qquad (9)$$

where $\mathcal{S}^N$ denotes the probability simplex with $N$ support points. Observe that for any $\alpha$, the set

$$\left\{x \in [0, \overline{\delta}] \mid j\epsilon - xT^\alpha(j) \geq k\epsilon - xT^\alpha(k) \ \forall k\right\}$$

is either empty or an interval and $\Psi_j$ is the probability of $\delta$ belonging to this interval under $F_\delta$. Since the time to execution $T^\alpha : \mathcal{S}^N \mapsto \mathbb{R}_+^N \bigcup \{\infty\}^N$ is continuous[2] function of $\alpha$ and $F_\delta$ is assumed to be continuous, $\Psi$ is a continuous function of $\alpha$. Thus, the Brouwer fixed point theorem implies that $\Psi$ has at least one fixed point. Since a fixed point of $\Psi$ is a solution to (8), an equilibrium satisfying (8) always exists.

If at an equilibrium $\alpha^*$, $T^{\alpha^*}$ is convex in $j$ then

$$h^*(j, \delta) = j\epsilon - \delta T^{\alpha^*} \tag{10}$$

is concave in $j$. Also, since $T^{\alpha^*}(j)$ is strictly increasing $j$, $h(j, \delta)$ is strictly decreasing difference in $(j, \delta)$. Thus, Theorem 10.6 in Sundaram (1996) implies that $\sigma^*(\delta) \in \arg\max h(j, \delta)$ is non-increasing in $\delta$. ∎

## 2.4 Numerical example

In the following example, we numerically compute the equilibrium in a market with elastic demand function.

Suppose $N = 50$, $\epsilon = 1$, $\delta \sim U[0, 160]$, $\lambda = 3$, $\mu = 12$ and the demand function be given by

$$\beta_j = \frac{1}{12}\left[0.5 + \left(\frac{N - j + 1}{15}\right)^2\right]$$

These parameters are chosen so as to get a full support equilibrium.

Theorem 1 guarantees us that

$$\alpha^* \in \min_{\alpha \in \mathcal{S}^N} \|\Psi(\alpha) - \alpha\|_2, \tag{11}$$

where $\Psi(\alpha)$ is defined in (9). We used MATLAB optimization routine FMINCON to solve optimization problem (11).

Figure 1 displays the expected execution time $T^{\alpha^*}$ and the supply thinning distribution $\alpha^*$. Figure 2 displays $\lambda$ times the equilibrium CDF of $\alpha^*$, i.e. the supply function, and the demand function $\mu\beta(p)$. For the choice of parameters in this example, the Walrasian market clearing price is approximately 28. Recall that the trading price available to a random buyer, i.e. the time average of the trading price, is distributed according to $\mathbb{P}(s^{\alpha^*} = j\epsilon | s^{\alpha^*} \leq N\epsilon) = \frac{\rho_j}{\sum_k \rho_k}$ corresponding to the tail CDF given in (6). For the parameters in this example, this distribution has a mean value of 23.77 and a standard deviation 15.55. Thus, the time-average of the prices is close to the Walrasian market clearing price. On the other hand, the mean and standard distribution of thinning distribution $\{\alpha_j^*\}$, i.e. the average price available to the sellers, are, respectively, 12.70 and 10.52.

---

[2] $T^\alpha$ is clearly continuous every $\alpha$ where it is finite and it approaches $\infty$ as $\sum_{j=1}^k \rho_k^\alpha$ approaches 1 for any $1 \leq k \leq N$ and hence is continuous at the boundary as well.

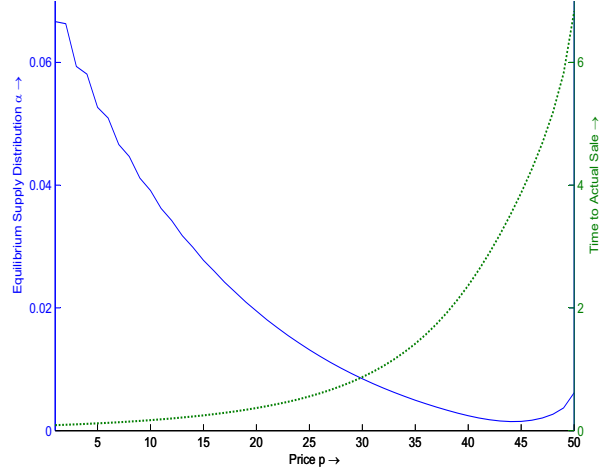

Figure 1: The equilibrium supply density and the execution time as a function of price

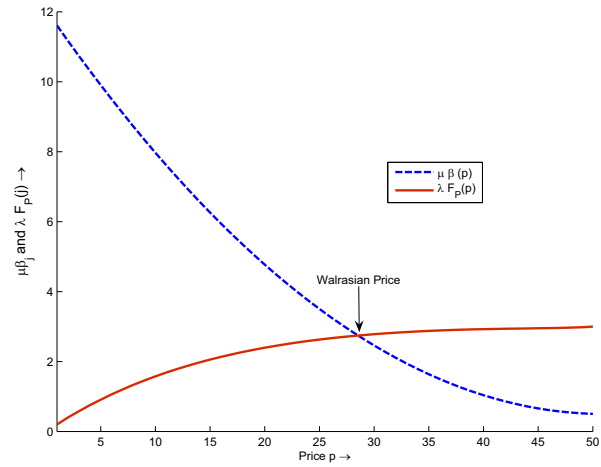

Figure 2: The equilibrium supply function and the demand function

This is significantly lower than the Walrasian price. A possible explanation for this phenomenon is that, since arriving sellers does not observe current outstanding price $s^{\alpha^*}$, they post a low price even though the current selling price is higher. Such low price orders are traded very fast, thus, does not contribute to the time averages.

Figure 2 plots the supply thinning function $\alpha_j^*$ and the demand decay $\beta_j$ as a function of the price-tick $j$. Notice that the equilibrium supply thinning function $\alpha_j^*$ decays significantly faster than the demand function $\beta_j$. This is because sellers not only want to reduce their waiting time by posting low prices but also faces competition in doing so from other sellers.

Figure 4 plots the expected inventory of outstanding sellers as a function of the price. The expected inventory is hump shaped as a function of price and increases rapidly at the right boundary because of the boundary effects, which is empirically observed in the context of limit order book in Bouchaud et al. (2002). We also observe that the shape of expected inventory conditional on current selling price is essentially independent of the current price. We expect this not to be the case if the sellers are allowed to condition their selling strategy on the current price because a seller facing a high current price would undercut more frequently than in this model resulting in smoother boundary in the density at the current price.

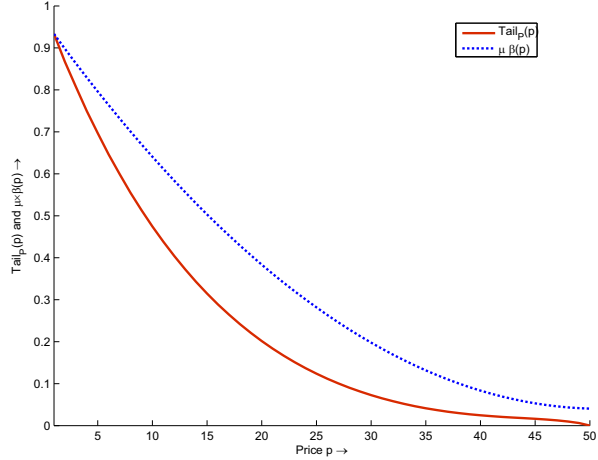

Figure 3: The $\beta_j$ and equilibrium Tail of $\alpha$ as a function of price.

## 3 Exchange with inelastic demand

In previous section, we observed that the demand elasticity has a serious impact on the equilibrium price distribution because sellers with low patience parameter are not able to force the buyers to pay a high price. In this section, we assume that the demand is inelastic in price, i.e. $\beta(p) = 1$ for all $p \in (0, \infty)$. Later in this section we show that an inelastic demand model is reasonable for most commodities in markets with very high trade frequency. Our main goal in this section is to investigate whether competition between sellers is sufficient to maintain reasonable (low) prices in a market with inelastic demand. The results in this section settle this question in the negative, i.e. sellers are able to leverage their market power to set very high prices. This result is consistent the result in Borenstein (2002) which finds that inelastic demand in conjunction with an continuous exchange mechanism gives rise to unprecedented high prices in the California electricity markets.

In this section, we work with continuous prices. We derive the differential equation characterizing the equilibrium thinning rate $\alpha^*(p)$ by taking appropriate limits in (8). Let $\epsilon \to 0$, $N \to \infty$, such that $N\epsilon \to \infty$ and the selling pricing grid

$$\{\epsilon, 2\epsilon, \ldots, N\epsilon\} \to (0, \infty).$$

Let $\alpha_P(p)$ and $F_P$ denote, respectively, the density and the CDF of the equilibrium selling price. Taking the limit in (10), the seller's expected surplus is given by

$$h(p, \delta) = p - \frac{\delta}{\mu} \left[ \frac{\rho F_P(p)}{(1 - \rho F_P(p))^2} + \frac{1}{(1 - \rho F_P(p))} \right] \qquad (12)$$

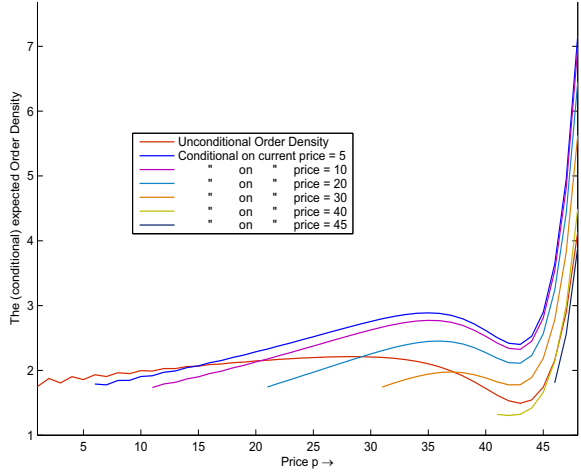

Figure 4: The (conditional) expected order densities

We focus on explicitly constructing equilibrium, and therefore, do not focus on whether (12) has a solution. After differentiating $h(p, \delta)$ with respect to $p$ and simplifying, the first order conditions are

$$\delta(p) = \frac{\mu}{2} \cdot \frac{(1 - \rho F_P(p))^3}{\rho \alpha_P(p)} \qquad (13)$$

Assume that $\sigma^*(\delta)$ is monotone in $\delta$. We later verify that in the equilibrium that we construct this indeed is the case.

The equilibrium conditions are equivalent to the following conditions on the tail of the selling price distribution $F_P$:

$$\mathbb{P}(\text{price } > p) = \mathbb{P}(\delta \leq \delta(p)) \quad \forall p \qquad (14)$$

In order to obtain a closed form solution, we assume that the patience parameter $\delta$ is distributed uniformly on $[0, \overline{\delta}]$. Thus, (13) is equivalent to the following implicit point-wise condition on the $F_P$:

$$\alpha_P(p) = \frac{d}{dp} F_P(p) = \frac{\mu}{2\overline{\delta}} \cdot \frac{(1 - \rho F_P(p))^3}{\rho(1 - F_P(p))} \qquad (15)$$

The following result follows from solving the above ordinary differential equation (ODE).

**Lemma 3.** *The equilibrium price distribution, i.e. the solution to the ODE (15), is given by*

$$F_P^*(p) = \begin{cases} 0, & p \leq 0, \\ \frac{\rho(1+\frac{\mu}{\overline{\delta}}p) - \sqrt{1-(1-\rho)(1+\rho(1+\frac{\mu}{\overline{\delta}}p))}}{\rho(1+\rho(1+\frac{\mu}{\overline{\delta}}p))}, & 0 \leq p \leq K, \\ 1, & p \geq K, \end{cases}$$

*where the support $K$ of the equilibrium selling price distribution is given by*

$$K = \frac{\overline{\delta}}{\mu} \cdot \frac{\rho}{1 - \rho}. \qquad (16)$$

As $\rho \to 1$, the support $K \to \infty$ and

$$lim_{\rho \to 1} F_P^*(p) = 1 - \frac{2\overline{\delta}}{2\overline{\delta} + \mu p}, \qquad p \geq 0. \tag{17}$$

Using (15) and (13) we get that the equilibrium selling strategy

$$\sigma^*(\delta) = F_P^{*-1}\left(1 - \frac{\delta}{\overline{\delta}}\right)$$

Since, $F_P^*$ is continuous on $[0, K]$ in Lemma 3 this equilibrium satisfy our assumption that $\sigma^*$ is monotone in $\delta$ over $[0, \overline{\delta}]$.

Lemma 3 establishes that as the congestion $\rho$ increases, the distribution of prices approach a power-law distribution, i.e. the equilibrium price distribution has very heavy tails even when the patience parameter $\delta$ is uniformly distributed between 0 and a finite upper bound $\overline{\delta}$. Thus, in congested markets with inelastic demand the sellers are able to leverage their market power to set (and obtain) very high prices. This partially explains the phenomena observed in California electricity markets Borenstein (2002).

The stationary expected inventory $Q(p)$ at price $p$ is given by

$$\begin{aligned} Q(p) &= \frac{\partial}{\partial p}\left\{\frac{\rho F_P(p)}{(1 - \rho F_P(p))}\right\}, \\ &= \frac{\mu}{2\overline{\delta}}\left(1 + \frac{(1-\rho)F_P(p)}{1 - F_P(p)}\right). \end{aligned} \tag{18}$$

We use Lemma 2 to compute the conditional expected density of outstanding orders. For all $p \geq s$, the conditional expected seller density

$$Q^c(p, s) \triangleq \mathbb{E}\left[Q(p)|\text{ current selling price} = s\right]$$

is given by

$$Q^c(p, s) = \frac{\partial}{\partial p}\left\{\frac{(\varrho_2(p) - \varrho_1)(1 - 2\varrho_1 + \varrho_1^2 + \varrho_1(\varrho_2(p) - \varrho_1))}{(1 - \varrho_1)^2(1 - \varrho_2(p))}\right\}$$

where $\varrho_2(p) \triangleq \rho F_P(p)$ and $\varrho_1 \triangleq \rho F_P(s)$. Simplifying this expression we obtain

$$Q^c(p, s) = \\ \left(\frac{1 - 3\varrho_1 + \varrho_1^2 + 2\varrho_1\varrho_2(p) - \varrho_1\varrho_2^2(p)}{(1 - \varrho_1)^2(1 - \varrho_2(p))^2}\right)\rho F_p(p) \tag{19}$$

## 3.1 Non-uniform patience distribution

Suppose the distribution $F_\delta$ of $\delta$ is given by

$$F_\delta(x) = \left(\frac{x}{\overline{\delta}}\right)^\gamma, \qquad 0 \leq \delta \leq \overline{\delta},$$

where $\gamma \in (\frac{1}{2}, 1]$. The ODE describing the CDF $F_P^\gamma(p)$ of equilibrium price distribution is given by

$$\frac{d}{dp}F_P^\gamma(p) = \frac{\mu}{2\overline{\delta}} \frac{(1 - \rho F_P^\gamma(p))^3}{\rho(1 - F_P^\gamma(p))^{\frac{1}{\gamma}}} \quad (20)$$

For $\rho = 1$, the equilibrium solution of the ODE (20) is given by

$$1 - F_P^{*\gamma}(p) = \frac{1}{\left(1 + \frac{(2\gamma-1)\mu}{2\gamma\overline{\delta}}p\right)^{\frac{\gamma}{2\gamma-1}}} \quad (21)$$

Using (18) the expected outstanding inventory is given by

$$Q_\gamma(p) = \frac{\mu}{2\overline{\delta}}\left(1 + \frac{(2\gamma-1)\mu}{2\gamma\overline{\delta}}p\right)^{\frac{1-\gamma}{2\gamma-1}} \quad (22)$$

By substituting $F_P^{*\gamma}$ from (21) in (13) and solving for $p$ in terms of $\delta$, we obtain that in equilibrium a seller with a patience parameter $\delta$ posts the price

$$p_\gamma^*(\delta) = \frac{2\gamma\overline{\delta}}{\mu(2\gamma-1)}\left(\left(\frac{\overline{\delta}}{\delta}\right)^{(2\gamma-1)} - 1\right) \quad (23)$$

At $\gamma = \frac{3}{4}$, we get a power tailed equilibrium selling price distribution as empirically observed in Farmer and Zovko (2002) in the context of limit order book. In this case, the outstanding order increases $Q(p) \sim \left(1 + \frac{p}{3}\right)^{\frac{1}{2}}$. Thus, the expected number of outstanding orders between 0 and $p$, i.e. the market depth,

$$D(p) = \int_0^p Q(p)dp \sim \left(1 + \frac{p}{3}\right)^{\frac{3}{2}}.$$

This market depth implies that the change in price $\frac{dp}{dQ}$ as a result of a market order of $Q$ units is of the order of $Q^{\frac{2}{3}}$, i.e. the price impact $\frac{dp}{dQ}$ is concave in the order quantity.

## 3.2 Summary of inelastic markets

The following observations follow form the discussion above.

(a) Everything else being equal, price is higher in less congested (low $\rho$) markets than in congested (high $\rho$) markets. This is because as $\rho$ increases, the time to execution at each price level goes up; consequently, the competition between sellers to obtain high prices becomes more intense.

(b) The equilibrium selling prices distribution exhibits power tails. This agrees with the empirical observations (see, e.g. Farmer and Zovko (2002)) in limit-book markets where the *market orders* are by definition not sensitive to prices. The market depth $D(p)$ and the price impact function prediction from this simple model agrees with observations in Daniels et al. (2003); Iori et al. (2003).

(c) From (21) it follows that as the market becomes congested, i.e. $\rho \to 1$, the equilibrium price distribution scales according to $p_s = \frac{\bar{\delta}}{\mu}$. Thus, when the patience parameter $\bar{\delta}$ is held constant, the equilibrium prices are high if $\mu$ is low, i.e. the market buy orders appear with a very low frequency; and vice versa.

As $\mu \to \infty$, i.e. the frequency of market buy orders increases, the price scaling $p_s \to 0$, i.e. the effective price window is very small. This has two implications: first, the sellers are not competing on price but on the executing time, and second, the assumption that the buyers are not price sensitive is not very serious.

(d) The distribution of the patience parameter $F_\delta$ has a significant impact on equilibrium market behavior.

## 4  Conclusion and Extensions

In our model a simple price vs waiting cost trade-off give rise to interesting market dynamics. The equilibrium in our model is numerically computable and can also be solved in closed form in special cases. The model predicts reasonable statistical behaviors of the online exchange market and also agrees with empirical data.

We believe that the equilibrium definition proposed in this paper can potentially be applied in other dynamic market clearing mechanisms. We are exploring the following extension.

(a) The observable quality of the commodity is heterogeneously distributed among the seller population. This results in multiple exchanges indexed by the quality traded and the buyers choosing the exchange to trade on based on their price-quality trade offs.

(b) Modeling order driven financial market where both sides of the market are patient and both buyers and sellers queue up.

(c) Service markets where buyer (sellers) are segmented based on attributes other than price and sellers (buyers) strategically assign resources to each segment.